\documentclass{emulateapj}
\usepackage{epsfig}
\usepackage{natbib}
\newcommand{\re}{R_e}
\newcommand{\sige}{\sigma_e}
\newcommand{\mstar}{M_*}
\newcommand{\mbh}{M_{\mathrm{BH}}}
\newcommand{\mdyn}{M_{\mathrm{dyn}}}
\newcommand{\fpp}{fundamental plane}

\shorttitle{Red Mergers and Massive Elliptical Galaxies}
\shortauthors{Boylan-Kolchin, Ma, \& Quataert}

\begin{document}

\title{Red Mergers and the Assembly of Massive Elliptical Galaxies:
  the Fundamental Plane and its Projections}
\author{Michael Boylan-Kolchin\altaffilmark{1}, Chung-Pei
  Ma\altaffilmark{2}, and Eliot Quataert\altaffilmark{2}}
\altaffiltext{1}{Department of Physics, University of California,
  Berkeley, CA 94720; mrbk@berkeley.edu} 
\altaffiltext{2}{Department
  of Astronomy, University of California, Berkeley, CA 94720}
\begin{abstract}
  
  Several recent observations suggest that gas-poor (dissipationless)
  mergers of elliptical galaxies contribute significantly to the
  build-up of the massive end of the red sequence at $z \la 1$.  We
  perform a series of major merger simulations to investigate the
  spatial and velocity structure of the remnants of such mergers.
  Regardless of orbital energy or angular momentum, we find that the
  stellar remnants lie on the \fpp\ defined by their progenitors, a
  result of virial equilibrium with a small tilt due to an increasing
  central dark matter fraction.  However, the locations of merger
  remnants in the projections of the \fpp\ -- the Faber-Jackson and
  $\re-\mstar$ relations -- depend strongly on the merger orbit, and
  the relations steepen significantly from the canonical scalings
  ($L\propto \sige^4$ and $\re\propto \mstar^{0.6}$) for mergers on
  radial orbits.  This steepening arises because stellar bulges on
  orbits with lower angular momentum lose less energy via dynamical
  friction on the dark matter halos than do bulges on orbits with
  substantial angular momentum.  This results in a less tightly bound
  remnant bulge with a smaller velocity dispersion and a larger
  effective radius.  Our results imply that the projections of the
  \fpp\ -- but not necessarily the plane itself -- provide a powerful
  way of investigating the assembly history of massive elliptical
  galaxies, including the brightest cluster galaxies at or near the
  centers of galaxy clusters.  We argue that most massive ellipticals
  are formed by anisotropic merging and that their \fpp\ projections
  should thus differ noticeably from those of lower mass ellipticals
  even though they should lie on the same \fpp.  Current observations
  are consistent with this conclusion.  The steepening in the
  $L-\sige$ relation for luminous ellipticals may also be reflected in
  a corresponding steepening in the $\mbh-\sige$ relation for massive
  black holes.
\end{abstract}


\section{Introduction}
Several independent lines of evidence point to the importance of
gas-poor (dissipationless) mergers in the assembly of massive
elliptical galaxies.  For example, the decreasing rotational support
and transition from disky to boxy isophotal shapes with increasing
stellar mass in elliptical galaxies suggests an increasing fraction of
dissipationless mergers in the growth of the most massive elliptical
galaxies (e.g., \citealt{bender1992, kormendy1996, faber1997,
  naab2006}).  The lack of correlation between metallicity
\citep{gallazzi2005} or metallicity gradients \citep{carollo1993} and
galaxy mass for massive ellipticals, while both are positively
correlated with galaxy mass for lower-mass ellipticals, provides
additional support for this picture.  There is also indirect evidence
from the evolution of galaxy luminosity functions
\citep{bell2004,faber2005,blanton2005a} and semi-analytic models of
galaxy formation \citep{kauffmann2000,khochfar2003, de-lucia2006} that
a significant fraction of low-redshift ellipticals were assembled via
dissipationless mergers.  Moreover, there is direct observational
evidence for a substantial number of ``red'' mergers between
early-type galaxies in an intermediate redshift galaxy cluster
\citep{van-dokkum1999, tran2005} as well as in more local galaxies
\citep{bell2005, van-dokkum2005}.  In spite of the importance of
gas-poor mergers, however, the properties of these mergers and their
remnants are not fully understood.

If elliptical galaxies gain substantial mass via dissipationless
merging at $z\la 1$, this process could have a significant impact on
their properties.  The purpose of this paper is to study how such
mergers affect the global dynamical and kinematic structure of the
remnants, as ellipticals show a very tight correlation in the space
spanned by their half-light radii ($\re$), central velocity
dispersions ($\sige$), and central surface brightnesses ($I_e$) known
as the \fpp\ \citep{dressler1987, djorgovski1987}.  Relationships
between two of the \fpp\ variables, such as the Faber-Jackson relation
\citep{faber1976} and the stellar mass-size relation, may also serve
as important constraints for galaxy formation models (e.g.,
\citealt{mcintosh2005}).

This is particularly true for the most luminous elliptical galaxies,
including brightest cluster galaxies (BCGs) that populate the
exponential tails of the galaxy luminosity, mass, and velocity
dispersion functions and contain the most massive black holes in the
universe.  These galaxies are likely to have undergone multiple
generations of gas-poor mergers, as repeated merging during cluster
formation is a well-motivated mechanism for BCG formation
\citep{merritt1985,dubinski1998}.  In this model, anisotropic infall
along cosmological filaments introduces a preferential direction in
the merging process \citep{west1994}, in contrast to mergers of
galaxies in less overdense environments.  As discussed below, we
suspect that this special assembly history has implications for the
scaling relations of massive ellipticals: the preferentially radial
orbits of mergers that form the most luminous ellipticals may lead to
different Faber-Jackson and stellar mass-size relations for these
galaxies than for normal ellipticals.

There are, in fact, tantalizing observational hints of deviations from
the canonical elliptical galaxy scaling relations for BCGs and other
luminous ellipticals.  For example, \citet{oegerle1991} find that
while BCGs lie on the same \fpp\ as lower mass ellipticals, their
central velocity dispersions change very little with increasing
luminosity and their effective radii increase steeply with luminosity,
significantly more so than for normal ellipticals.  A similar effect
is present in the early-type galaxy sample of the Nuker team (see
fig.~4 of \citealt{faber1997} and Lauer et al. 2006, in prep.) as well
as in the \citet{postman1995} BCG sample \citep{graham1996a}, while
\citet{wyithe2006} has claimed to detect an analogous change in the
$\mbh-\sige$ relation for the most massive black holes.

In this paper, we present the results of a large number of simulations of
major mergers of elliptical galaxies.  Our galaxy models contain dark
matter haloes and stellar bulges, both simulated with sufficient force and
mass resolution to reliably investigate the remnants down to scales of
$\sim 0.1 \re$ (over $10^6$ total particles per simulation).  We consider a
wide variety of energies and angular momenta for the merger orbits and
quantify the effects of orbital parameters on the merger remnants.  This
work is a significant extension of our earlier related study
\citep[hereafter BMQ]{boylan-kolchin2005}, where we considered only a few
possible orbits for the merging galaxies.  The simulations presented in
this paper allow us to study in detail the effects of dissipationless
merging and merger orbits on the \fpp\ and its projections.
Section~\ref{sec:methods} describes the initial galaxy models and
simulation parameters.  Section~\ref{sec:results} contains the results of
our simulations, including the locations of merger remnants in the \fpp\
(Section~\ref{sec:fp}) and its projections (Section~\ref{sec:projections}).
In Section~\ref{sec:discuss} we discuss our results and their implications
for the assembly of massive elliptical galaxies and their central black
holes.

\section{Methods}
\label{sec:methods}

The galaxy models used in our simulation consist of dark matter haloes and
stellar spheroids and are fully described in BMQ; we review the essential
features here.  The initial dark matter haloes have the \citet{navarro1997}
density profiles with a virial mass $M_{200}=10^{12} \, M_{\odot}$ and a
concentration $c=10$, resulting in a virial radius of 162.6 kpc and scale
radius of 16.26 kpc.  The initial stellar bulges have the
\citet{hernquist1990} density profiles with a total stellar mass
$\mstar=5\times 10^{10} \,M_{\odot}$, resulting in a stellar baryon
fraction of $f_b=0.05$.  The additional free parameter in the stellar
profile, the scale radius $a$, is set by the $\re-\mstar$ relation measured
for SDSS elliptical galaxies by \citet{shen2003}:
\begin{equation}
\re=4.16 \left(\frac{\mstar}{10^{11}\,M_{\odot}} \right)^{0.56}
\, \mathrm{kpc},
\label{eqn:shen}
\end{equation}
where $\re=1.8153\,a$ for the Hernquist profile; this gives $\re=2.82$ kpc
for our initial models.  This relation is appropriate for low-redshift
elliptical galaxies.  Ideally, we would use the $\re-\mstar$ relation for
$z \approx 1$ rather than $z \approx 0$.  The data that are available at
higher redshift indicate there has not been strong evolution in the
$\re-\mstar$ relation from $z=1$ to $z=0$ \citep{trujillo2004a,
  mcintosh2005}.  We include the effects of baryonic dissipation via the
adiabatic contraction model of \citet{blumenthal1986}.  Although the
applicability of this model to elliptical galaxies is not fully obvious,
recent hydrodynamic simulations suggest that it is surprisingly accurate
\citep{gnedin2004}.  The resulting (circular) aperture velocity dispersion
within $\re$ for the stellar component in our models is $\sige=151$ km/s.
We also perform a number of unequal mass mergers, with mass ratio 0.33:1,
where the less massive galaxy also has $f_b=0.05$ and a scale radius set by
equation~(\ref{eqn:shen}).  Although we present the parameters of our
initial conditions in physical units, the simulations reported here can be
readily scaled to other stellar and dark matter masses (see associated
discussion in Section~\ref{sec:projections}).

We use {\tt GADGET-1} \citep{springel2001} to perform the merger
simulations.  All runs use a force softening of 0.3 kpc ($=0.106 \re$)
and equal-mass particles for the stellar and dark matter components:
$N_*=2.5 \times 10^4$, $N_{\mathrm{DM}} = 5\times 10^5$ for each
galaxy model.  The initial particle positions of each component are
sampled from the component's spherically symmetric density profile.
To ensure that the models are in equilibrium, we compute separate
distribution functions for the halo and bulge and then use these
distribution functions to initialize the particle velocities.  Our
initial galaxy models do not include any velocity anisotropy.  As
\citet{nipoti2002} show, however, the inclusion of a modest amount of
velocity anisotropy in models of merging galaxies does not
substantially change the properties of the remnants, while adding
significant velocity anisotropy makes the initial model unstable.
Additionally, \citet{ciotti1996} show that velocity anisotropy cannot
be the primary origin for the tilt in the \fpp.

$N$-body models set up with this procedure are in general extremely
stable over many dynamical times (e.g., \citealt{kazantzidis2004}),
which we have explicitly checked for our models.  An additional source
of spurious evolution inherent to $N$-body simulations is two-body
relaxation, a consequence of using simulation particles that are many
orders of magnitude more massive than the stars or dark matter
particles they represent.  We have performed test runs to ensure that
two-body relaxation does not affect our results for the duration of
the simulations ($\sim 6$ Gyr) for our choices of particle number and
force resolution (see BMQ for detailed resolution and stability
studies).

\begin{deluxetable}{cclccc}
\tablecaption{Summary of Merger Simulations
\label{table-ICs}}
\tablehead{
\colhead{\textbf{Run}\tablenotemark{a}}
& \colhead{\textbf{Mass}\tablenotemark{b}} 
& \colhead{\textbf{Orbit}\tablenotemark{c}} 
& \colhead{$\mathbf{V_{orb}}$\tablenotemark{d}} 
& \colhead{$\mathbf{r_p}$\tablenotemark{e}} 
& \colhead{\boldmath{$\epsilon$}\tablenotemark{f}}
}
\startdata
P1 & 1:1 & Parabolic & 250  & 0    & --   \\    
P2 & 1:1 & Parabolic & 250  & 2.29 & --   \\    
P3 & 1:1 & Parabolic & 250  & 12.5 & --   \\    
P4 & 1:1 & Parabolic & 250  & 37.8 & --   \\    
P5 & 0.33:1 & Parabolic & 222  & 0    & --  \\
P6 & 0.33:1 & Parabolic & 222  & 2.31 & --  \\
P7 & 0.33:1 & Parabolic & 222  & 12.5 & --  \\
P8 & 0.33:1 & Parabolic & 222  & 20.5 & --  \\
B1 & 1:1 & Bound & 200  & 0    & 0.00 \\    
B2 & 1:1 & Bound & 200  & 2.29 & 0.10 \\    
B3 & 1:1 & Bound & 200  & 20.8 & 0.30 \\    
B4 & 1:1 & Bound & 200  & 37.8 & 0.40 \\    
B5 & 1:1 & Bound & 200  & 57.1 & 0.48 \\    
B6 & 1:1 & Bound & 200  & 78.4 & 0.56 \\    
\enddata
\tablenotetext{a}{Name of run}
\tablenotetext{b}{Mass ratio of progenitors}
\tablenotetext{c}{Energy of run (either parabolic or bound)}
\tablenotetext{d}{Initial orbital velocity (km s$^{-1}$) for  
     a galaxy with $M_{\mathrm{dm}}=10^{12} M_\odot$ and $M_*=5\times
     10^{10} M_\odot$}
\tablenotetext{e}{Pericentric distance of orbit (kpc)}
\tablenotetext{f}{Circularity of orbit (undefined for parabolic orbits)}
\end{deluxetable}

The centers of the galaxies in the merger simulations are initially
separated by the combined virial radii of the two haloes and the
orbital parameters are defined in the standard two-body point mass
approximation (\citealt{binney1987}, Appendix 1D;
\citealt{khochfar2006}).  In this limit, orbits can be characterized
by two quantities: energy $E$ and angular momentum $L$, or
equivalently, pericentric distance $r_p$ and eccentricity $e$ (or
circularity $\epsilon \equiv \sqrt{1-e^2}$ for runs with $E<0$).  The
orbital parameters for all of our production runs are listed in
Table~\ref{table-ICs}.  Since our goal is to understand the effects of
different orbits on the properties of the merger remnants, we consider
both parabolic encounters and bound orbits, and for each energy we
perform simulations over a range of orbital angular momenta.  The
energy of the bound orbits is taken to correspond to the most probable
energy from \citet{benson2005}, which is based on a study of dark
matter halo mergers averaged over all cosmological environments.  We
use this value of $E$ only as a rough guide and note that existing
statistical studies of dark matter halo mergers are dominated by low
mass objects and therefore do not necessarily represent the orbits
responsible for the assembly of massive elliptical galaxies in cluster
environments.  Larger samples of haloes should soon provide robust
statistical information about the environmental dependence of halo
mergers.  In particular, as we argue later in this paper, major
mergers leading to the formation of central massive galaxies in
clusters may follow preferentially radial orbits.

The non-spherical nature of the luminous portion of merger remnants
means that quantities such as $\re$ and $\sige$ depend on the
orientation at which the objects are viewed.  An immediate consequence
is that an intrinsic scatter in the \fpp\ projections will exist
solely due to a variation in viewing angle.  In order to compute $\re$
and $\sige$ that can be compared with observations, and to quantify
the effects of viewing angle, we observe the remnant in each
simulation from $10^4$ different lines of sight chosen randomly from
points on a sphere.  The effective radius is determined by the radius
of the circle containing half of the projected stellar mass and the
mass-weighted aperture velocity dispersion is computed within one
$\re$.

\section{Results}
\label{sec:results}

\subsection{The Fundamental Plane}
\label{sec:fp}

The \fpp\ is typically expressed in the coordinate system
defined by the observables $(x,y,z)=(\log \sige, \log I_e, \log \re)$ as
\begin{equation}
  \re \propto \sige^a I_e^{-b} \,,\quad {\rm or} \quad
    z=ax-by+z_0  \,,
\label{plane1}
\end{equation}
where $a$ represents the angle of the plane from the vertical
axis $z=\log \re$, and $b/a$ represents the amount of rotation about
the $z$-axis.  If we write 
\begin{equation}
\mstar(<\re) \equiv c_1 \, I_e \re^2 \quad \mathrm{and}  \quad
\mdyn(<\re) \equiv c_2  \, \frac{\sige^2 \re} {G}\,,
\label{eqn:constants}
\end{equation}
where $c_1$ and $c_2$ are structure coefficients, then the \fpp\
observables are related by
\begin{equation}
\re \propto \frac{c_2}{c_1} \left(\frac{\mdyn}{\mstar}\right)^{-1}
\sige^2 I_e^{-1} 
\end{equation}
\citep{bender1992}.  Therefore, if elliptical galaxies form a
homologous sequence (if $c_1$ and $c_2$ are constant) with light
tracing dynamical mass and a constant stellar mass-to-light ratio,
they should lie on the virial plane given by $a=2,\,b=1$.  However,
the observed \fpp\ differs from the virial plane.  For example,
$a=1.24 \pm 0.07$, $b=0.82 \pm 0.02$ in the Gunn r-band
\citep{jorgensen1996} and $a=1.53 \pm 0.08$, $b= 0.79 \pm 0.03$ in the
K-band \citep*{pahre1998a}.  Results for all SDSS wavebands are similar
to the K-band \fpp\ \citep{bernardi2003a}.  This difference represents
a ``tilt'' in the observed \fpp\ from the virial plane: 
either $c_2/c_1$, $\mdyn/\mstar$, or $\mstar/L$ varies systematically
with galaxy mass.

We are able to investigate the first two possibilities with our
simulations, but we cannot directly test the third.  In what follows,
we assume a constant stellar mass-to-light ratio for the galaxies in
our simulations.  We can therefore interpret our results as testing
the amount of the \fpp\ tilt that is due to either non-homology or
central dark matter fraction.  A number of authors have argued that
stellar population variations with galaxy mass are responsible for the
\fpp\ tilt (e.g.  \citealt{gerhard2001, treu2005, cappellari2006}).
As \citet*{pahre1998} point out, however, the existence of a
significant tilt in the near-infrared, where stellar population
effects should be less pronounced than in the optical, indicates that
stellar populations alone are not the cause of the \fpp\ tilt.
Furthermore, K-band $\mstar/L$ ratios do not vary strongly with the
age of the stellar population for relatively old stellar populations
that are appropriate for the galaxies we are simulating.  Analyses of
SDSS early-type galaxies \citep{kauffmann2003, padmanabhan2004} also
find that stellar mass-to-light ratios are relatively constant with
mass.  We therefore believe that assuming a constant stellar
mass-to-light ratio is a reasonable approximation.


\begin{figure*}
\begin{center}
\includegraphics[scale=0.4]{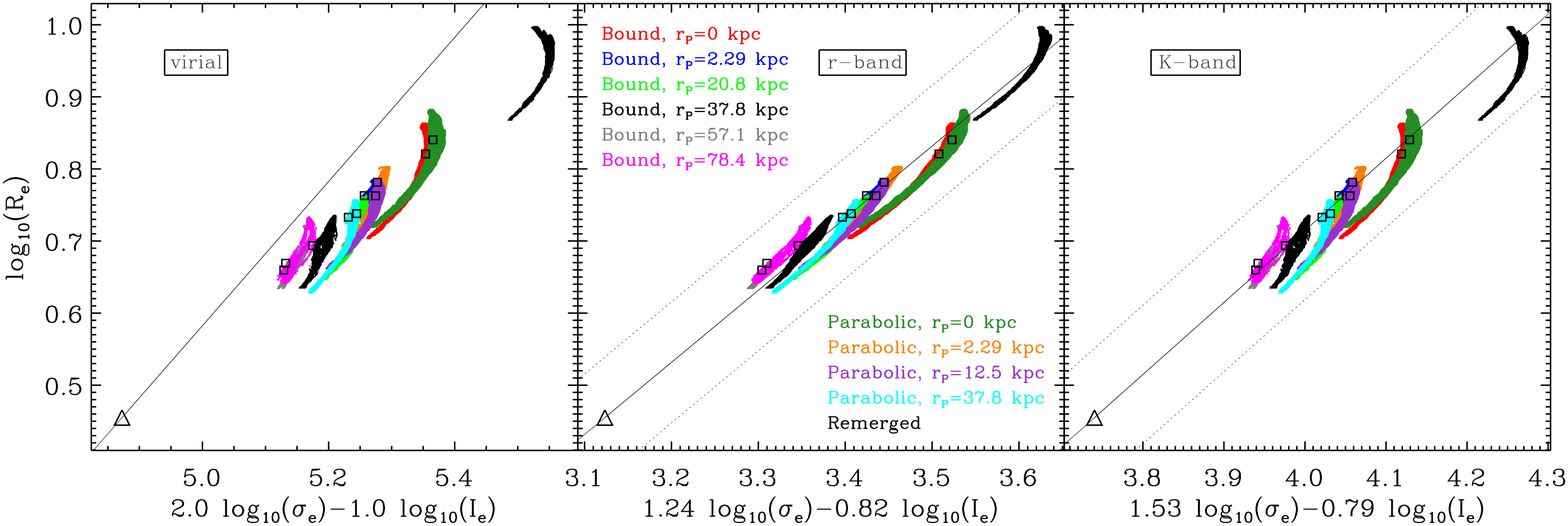}
\caption{Locations of our equal-mass merger remnants in the plane in
  $(\log\sige,\log I_e,\log \re)$-space that corresponds to the
  edge-on projection of the virial plane (left), r-band \fpp\ (middle;
  \citealt{jorgensen1996}), and K-band \fpp\ (right;
  \citealt{pahre1998a}).  The solid line in each panel indicates a
  slope of 1 for the edge-on view; the dotted lines in the middle and
  right panels indicate the observed $1-\sigma$ scatter in the
  direction of $\re$.  For each panel, each simulation (corresponding
  to a particular energy and angular momentum) is plotted with a
  different colour.  Each point of a given colour corresponds to one
  of $10^4$ viewing angles, while the square marks the ``most
  probable'' remnant.  A single re-merger simulation, using the
  remnants of runs P3 and B6 as initial galaxies and with a parabolic
  orbit with $r_p=21$ kpc, is plotted in black in the upper right
  corner as well.  The initial galaxy model is a single point (large
  triangle).  The effective radii are measured in kpc, the velocity
  dispersions in km s$^{-1}$, and surface brightnesses in $10^{10}\,
  M_{\odot} \,\mathrm{kpc}^{-2}$ for all panels.}
\label{fig:fp}
\end{center}
\end{figure*}

\begin{figure*}
\begin{center}
\includegraphics[scale=0.7]{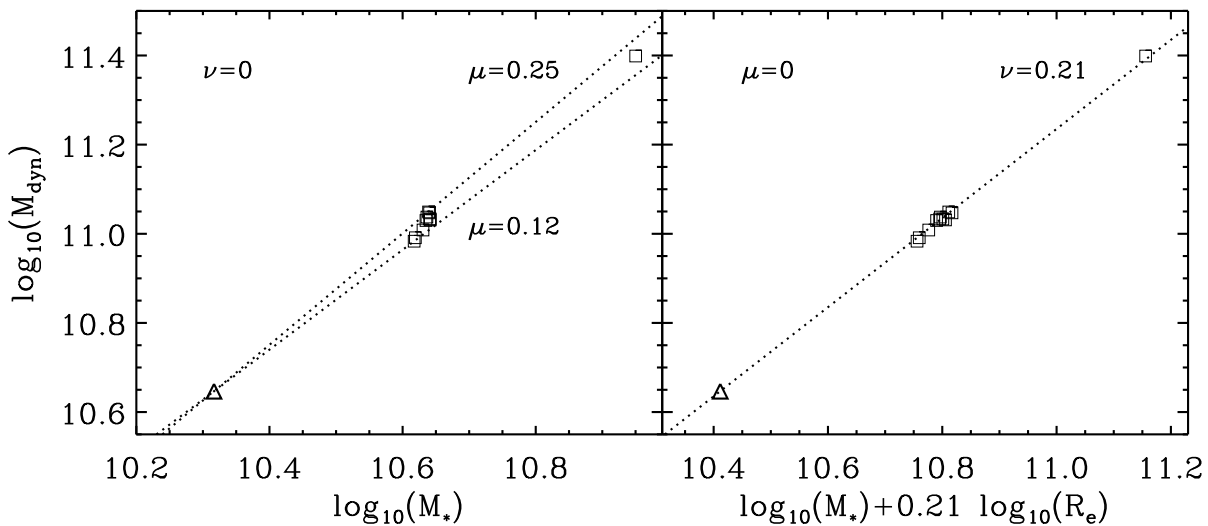}
\caption{
  Locations of the merger remnants from Fig.~\ref{fig:fp} in
  $(\log\mdyn,\log\mstar,\log \re)$-space: $\mdyn \propto \mstar
  ^{1+\mu} \re^{\nu}$.  Both $M_{\mathrm{dyn}}$ and $\mstar$ are
  computed within a sphere of radius $\re$.  {\it Left:} The $\nu=0$
  plane, which is a good approximation to the r-band plane (fig.~1b).
  The dotted lines correspond to $M_{\mathrm{dyn}} \propto
  \mstar^{1+\mu}$ with $\mu=0.12$ (lower) and $\mu=0.25$ (upper);
  these bracket the results of our simulations.  {\it Right:} The
  $\mu=0$ plane, which fits the observed K-band \fpp\ well.  We plot
  the parameterization of the tilt in the \fpp\ that best fits our
  simulations: $\mdyn/\mstar \propto \re^{0.21}$ (dotted line).  The
  effective radii are measured in kpc, while the stellar and dynamical
  masses are measured in $M_{\odot}$.}
\label{fig:tilt}
\end{center}
\end{figure*}

Fig.~\ref{fig:fp} shows the locations of all of our equal-mass merger
remnants in the edge-on projection of each of the three planes discussed
above -- virial, r-band, and K-band.  We find that the 0.33:1 merger
remnants show all of the same trends as the 1:1 remnants and thus the
former are not plotted here.  The $1-\sigma$ errors of the \fpp\ in the
direction of $\re$ are indicated by the dotted lines in Fig. \ref{fig:fp}.
Each remnant is marked with a different colour, and each point represents
one random viewing angle of the remnant.  The initial galaxy is spherical
by construction, so it is located at a single point in the \fpp\ (large
triangle; lower-left corner).  The stellar mass-to-light ratio has been
chosen so that the initial model lies on the observed \fpp\ (this
corresponds to K-band and r-band mass-to-light ratios of 1.0 and 3.6,
respectively); the same stellar mass-to-light ratios are used to place the
merger remnants on the \fpp.

The left panel of Fig.~\ref{fig:fp} shows that all of the remnants lie
significantly off the simple virial theorem expectation.  Strikingly,
however, \emph{all} of the simulated observations of \emph{all} of the
merger remnants lie well within the $1-\sigma$ errors of the \fpp\ for
both the optical and near-infrared planes.  These results establish
that the remnants of major mergers of spherical elliptical galaxies
lie on the fundamental plane defined by their progenitors for the wide
range of orbital energies and angular momenta covered by our
simulations.  We note, however, that our dynamic range in $R_e$ and
$M_*$ is rather modest, and all of our simulations begin at the same
point in the \fpp.  Thus our simulations cover only a limited part of
the observed plane.

Since massive elliptical galaxies may have undergone multiple
dissipationless mergers during their assembly history, we also
performed a simulation in which the two progenitors were randomly
chosen to be the remnants of runs P3 and B6 and the orbit was
parabolic with $r_p=21$ kpc.  We find this re-merger remnant lies on
the same \fpp\ as the initial galaxy model and first-generation merger
remnants (Fig.~\ref{fig:fp}, black region in upper right corner).
This single re-merger simulation is certainly not definitive, but it
does support the hypothesis that the presence of elliptical galaxies
on the \fpp\ is consistent with multiple generations of
dissipationless merging.

While the observed \fpp\ is typically plotted in the $(x,y,z)=(\log
\sige, \log I_e, \log \re)$ space as shown in Fig.~1, we find it
useful to interpret the plane theoretically in the transformed
coordinate system $(x',y',z')=(\log \mdyn, \log \mstar, \log \re)$, in
which a plane can be written as
\begin{equation}
     \mdyn \propto \mstar^{1+\mu} \re^{\nu} \,,\quad {\rm or}\quad
     z' = {1\over \nu} \left[ x' - (1+\mu) y'\right] + z'_0\,,
\label{plane2}
\end{equation} 
where both $\mdyn$ and $\mstar$ refer to the masses enclosed within
$\re$.  The parameters $\mu$ and $\nu$ quantify how the ratio of dark
matter mass to stellar mass within $\re$ varies with increasing galaxy
mass and size.  

Comparing equations~(\ref{plane1}) and (\ref{plane2}), and using the
definitions in equation~(\ref{eqn:constants}), gives
\begin{equation}
       \mu={2b\over a}-1 \,, \quad \nu=1+ {2(1-2b)\over a} \,.
\end{equation}
The virial plane of $a=2,\,b=1$ corresponds to $\mu=0,\, \nu=0$ (i.e.
$\mdyn \propto \mstar$); it is therefore conveniently parallel to axis
$z'=\log \re$ in the $\mdyn-\mstar-\re$ space.  The observed r-band
and K-band fundamental planes have $(\mu,\nu)=(0.323,-0.0323)$ and
$(\mu,\nu)=(0.0327,0.242)$, respectively.  It is interesting to note
that the r-band plane of \cite{jorgensen1996} has $\nu\approx 0$, i.e.
$\mdyn/\mstar \propto \mstar^\mu$, while the K-band \fpp\ has
$\mu\approx 0$, i.e.  $\mdyn/\mstar \propto \re^\nu$.

Fig.~2 shows the locations of our merger remnants in the edge-on view
of the $\nu=0$ (left panel; appropriate for r-band) and $\mu=0$ (right
panel; appropriate for K-band) planes.  The remnants appear more
consistent with the K-band plane than the r-band plane.  Although
Fig.~2 and Fig.~1bc display the same information in different
coordinate systems, Fig.~2 illustrates clearly how the dynamical mass
within $\re$ increases more than the stellar mass within the same
radius, meaning that the inner regions of our merger remnants are more
dark matter-dominated than their progenitors.  This increasing dark
matter fraction with increasing galaxy mass is the primary cause for
the tilt of the observed \fpp\ from the virial plane in our gas-free
mergers.

Since systematic non-homology and varying central dark matter fraction
in elliptical galaxies are often considered as possible origins for
the tilt of the \fpp\ (e.g., \citealt{faber1987, ciotti1996,
  graham1997, pahre1998, gerhard2001, padmanabhan2004, trujillo2004}),
it is useful to assess the importance of non-homology in the context
of the two-component dark matter and stellar models considered in this
paper.  We find that both $c_1$ and $c_2$ [as defined in
equation~(\ref{eqn:constants})] are constant for our merger
remnants\footnote{Note that we compute the surface \emph{mass}
  density, while it is the surface \emph{luminosity} density that is
  measured observationally.  The two are simply related by the stellar
  mass-to-light ratio, which we assume is constant.}: $c_1=2.70$ with
a maximum deviation of $<4\%$ and $c_2=2.90$ with a maximum deviation
of $<3\%$.  This implies that the final systems (bulge+halo) are in
virial equilibrium and are dynamically homologous (because $c_2$ is
constant) and that the luminous portions of our remnants are
structurally homologous (because $c_1$ is constant).  The total system
is structurally non-homologous, however, due to a change in the dark
matter halo relative to the bulge.  The inner part of the halo, which
contains the bulge, does not grow in radius as much as the bulge does
during the course of each merger, while the outer part of the halo
expands significantly.  This results in an increased central dark
matter fraction for the remnants, as shown in Fig.~2.  The amount of
tilt in our simulations is very similar to the observed \fpp\ tilt,
implying that there is no need to invoke stellar mass-to-light
variations to account for the tilt in our simulations.

As the bulges themselves \emph{are} structurally homologous ($c_1$ is
constant and the density profiles of the remnant bulges are well-fit
by a Hernquist profile), we refer to a varying dark matter fraction
rather than a structural non-homology as the origin of the tilt.  We
note that \citet{graham1996} find S\'{e}rsic $r^{1/n}$ models provide
better fits to BCG surface brightness profiles than the $r^{1/4}$ law
and that $n$ correlates positively with mass, indicating a real
non-homology in the elliptical galaxy population that is not captured
by our simulations (but see \citealt{gonzalez2005} for an argument
that two separate $r^{1/4}$ profiles fit BCGs significantly better
than S\'{e}rsic profiles).  

A further point of interest is that the scatter due to viewing angle
variation of the non-spherical remnants in our simulations
(represented by the spread within each colour in Fig.~\ref{fig:fp})
is roughly parallel to the observed fundamental plane.  This trend is
a consequence of the virial equilibrium of individual galaxies: at
fixed \emph{stellar} mass, $\sige^2 \re \sim \re^{0-0.33}$, so $\re
\propto \sige^{-(2-3)}$.  Intuitively, a non-spherical remnant
maintains a small line-of-sight velocity dispersion by extending in
the perpendicular direction.  This shows that while viewing angle
variation introduces some scatter into the \fpp\ 
\citep{gonzalez-garcia2003}, this scatter tends to be along the \fpp\ 
itself.  In addition, the scatter due to viewing angle variation does
not increase between the first and second generation mergers, further
indicating that the \fpp\ scatter is relatively insensitive to
dissipationless merger history.

Similar conclusions about mergers and remergers preserving the \fpp\ 
have been reached in various settings by previous authors
\citep{capelato1995, nipoti2003, gonzalez-garcia2003, robertson2005}.
The previous remerger simulations most relevant to this work are those
of \citet{nipoti2003} and \citet{robertson2005}.  They show that both
remergers of two-component galaxies similar to ours and remergers of
triaxial remnants of gaseous disk galaxy mergers lie on the (K-band)
\fpp.  These results suggest that the spherical symmetry typically
assumed in initial galaxy models of merger simulations, albeit
unrealistic, does not significantly affect the results on the
preservation of the \fpp.  Exploring the evolution of the \fpp\ under
dissipationless merging for general triaxial initial conditions would
be useful; however, we believe that a brute-forced sampling of the
large parameter space associated with merging triaxial galaxies is
highly inefficient.  We will discuss a more economical procedure in
Section 4 that uses the results of cosmological simulations to
motivate realistic merging sequences.

\subsection{Projections of the Fundamental Plane}
\label{sec:projections}

The fact that our merger remnants are consistent with the \fpp\ does
not necessarily imply that the properties of the remnant are
insensitive to the merger orbit, nor does it imply that the remnants
necessarily follow the observed Faber-Jackson or $R_e-M_*$ relations:
requiring a galaxy of a given mass to lie on the mean of both \fpp\ 
projections picks out a single point in $\re-\sige-I_e$ space, whereas
the \fpp\ at fixed mass forms a line in the same space (an $\re-\sige$
correlation).

\begin{figure*}
\begin{center}
\includegraphics[scale=0.9]{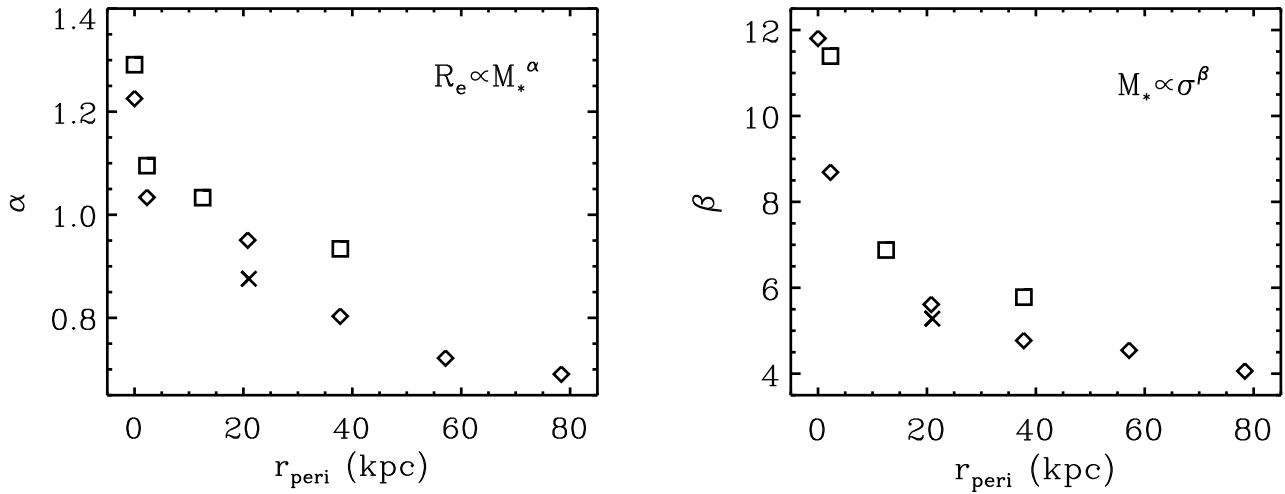}
\caption{
  Logarithmic slopes of the mass-size relation ($\re\propto
  \mstar^\alpha$; left) and Faber-Jackson ($\mstar\propto
  \sige^\beta$; right) relations as a function of orbital pericentric
  distance for each of the equal-mass merger simulations listed in
  Table~\ref{table-ICs}.  The two orbital energies are shown as
  diamonds (bound) and squares (parabolic), while the re-merger run is
  marked with an X symbol.  At a given $r_{\rm peri}$, the bound
  orbits produce remnants with a slightly lower $\alpha$ and $\beta$
  than parabolic orbits.  The parabolic head-on orbit has $\beta
  \approx 28$, so it is not plotted here.  For orbits with significant
  orbital angular momentum (large $r_{\rm peri}$), our merger
  simulations reproduce the observed values of $\alpha\sim 0.6$ and
  $\beta\sim 4$ but we predict a sharp increase in $\alpha$ and
  $\beta$ for more radial mergers.  The results for the re-merger run
  agree well with those for the first generation mergers, indicating
  that the spherical and isotropic galaxy models used in the first
  generation mergers do not strongly bias our results.}
\label{fig:rperi-slopes}
\end{center}
\end{figure*}

In order to quantify the effects of merger orbits on the projections
of the \fpp, we compute the slopes in the $\re \propto
\mstar^{\alpha}$ and $\mstar \propto \sige^{\beta}$ relations for each
of our 1:1 merger simulations.  Fig.~\ref{fig:rperi-slopes} shows the
dependence of the slopes $\alpha$ (left) and $\beta$ (right) as a
function of the pericentric distance of the orbit\footnote{At fixed
  energy, varying the angular momentum is equivalent to changing the
  pericentric distance.} for both the bound orbits (diamonds) and
parabolic orbits (squares).  The slopes have a strong trend with
pericentric distance: increasing $r_p$ corresponds to decreasing
$\alpha$ and decreasing $\beta$.  These trends were already noted in
BMQ based on a few orbits, but the comprehensive set of simulations
with different $L$ and $E$ presented in this paper provide a much more
quantitative understanding of the impact of the merger orbit on the
\fpp\ projections.  The slopes for the re-merger remnant, marked with
an X in Fig.~\ref{fig:rperi-slopes}, agree well with those for the
first generation mergers.  Although Fig.~\ref{fig:rperi-slopes} shows
only the results of 1:1 merger simulations, we find similar trends in
0.33:1 mergers.  These results suggest that the angular momentum and
energy of the orbit have the biggest impact on the \fpp\ projections,
with the mass ratio of the merging galaxies and the internal structure
of the initial bulge/halo system being of secondary importance.

For comparison, the observed \fpp\ projections are $\re \propto
L^{0.56-0.63}$ and $L \propto \sige^{4-4.14}$
\citep{pahre1998a,bernardi2003}, although as noted in Section 1, these
scalings are probably not applicable to the most luminous ellipticals.
Our high angular momentum simulations have scaling relations in good
agreement with both of the observed projections.  The low angular
momentum runs, however, deviate significantly from these canonical
scalings (similar deviations for low angular momentum runs were also
reported in \citealt{nipoti2003}, BMQ, and \citealt{robertson2005}).

The combination of Figs.~\ref{fig:fp} and \ref{fig:rperi-slopes}
reveals that while all of the remnants lie on the \fpp, the remnant
properties are nonetheless strongly affected by merger orbits.
Differences in merger orbits can lead to significant differences in
the projected scaling relations, but variations in the $\re-\mstar$
relation are compensated by corresponding variations in the
$\mstar-\sige$ relation, maintaining the \fpp\ because of virial
equilibrium of the bulge-halo system.

It is important to point out that although we have quoted our results
in terms of galaxies with specific masses and radii, they can be
easily rescaled to other masses and radii.  Such a rescaling will
affect the amplitude of the relations we are studying but not the
slopes.  For example, if we increase all masses by a factor of 10 and
all radii by a factor of $10^{1/3}$ (to preserve virial scalings), the
velocity dispersion will scale as $\sqrt{M/R}=10^{1/3}$.  This moves
our initial conditions off of the $\re-\mstar$ relation, since
$\re=2.83*10^{1/3}=6.1$ kpc rather than the 10.2 kpc given by the Shen
et al. relation.  This difference is comparable to the observed
scatter in the relation, $\sigma_{\ln r} \approx 0.3$
\citep{shen2003}.  The velocity dispersion will also differ from that
predicted by the Faber-Jackson relation: $\sige=151*10^{1/3}=325$
km s$^{-1}$ rather than $151*10^{1/4}=269$ km s$^{-1}$.  Thus mergers
of galaxies at these scales have remnants with identical $\re-\mstar$
and $\mstar-\sige$ slopes as those given above but with different
amplitudes than the means of the observed relations.  The slopes of
the relations quoted here are therefore quite robust.

Qualitatively, the dramatic steepening of the $\re-\mstar$ and
$\mstar-\sige$ relations with decreasing orbital angular momentum in
Fig.~\ref{fig:rperi-slopes} can be understood by considering the role
of dynamical friction and energy transfer during a merger.  At a given
orbital energy, stellar bulges with substantial angular momentum need
to lose more angular momentum (and accordingly, more energy) via
dynamical friction on the background dark matter halo than do bulges
on low angular momentum orbits.  This leads to a more tightly bound
remnant bulge with a larger $\sigma$ (smaller $\beta$) and smaller
$R_e$ (smaller $\alpha$), as is found in the simulations.  The same
effect also explains why orbits with more initial orbital energy (our
parabolic orbits) tend to lead to less bound remnants with smaller
$\sigma$ (larger $\beta$) and larger $R_e$ (larger $\alpha$).

Quantitatively, we can understand the essence of the physics involved
in producing the trends seen in Fig.~\ref{fig:rperi-slopes} by
examining the energy equation for the stellar bulges.  The energy
conservation equation for two identical bulges, each with initial
stellar mass $M_{*,i}$ and effective radius $R_i$, merging to form a
final bulge of mass $M_{*,f}$ and effective radius $R_f$ can be
written as
\begin{equation}
f_f \frac{M_{*,f}^2}{R_f} = 2 f_i \frac{M_{*,i}^2}{R_i}+ \eta
\frac{M_{*,i}^2}{2R_i} \,.
\label{eqn:homo1}
\end{equation}
The parameter $f$ depends on the structure of the
bulge and surrounding dark matter halo:
\begin{equation}
f \frac{\mstar^2}{\re} \equiv \frac{1}{2} 
\int \frac{\rho_*(r) \mdyn(r)}{r} \mathrm{d^3 x} \,,
\label{eqn:f}
\end{equation}
where $\mdyn(r)=\mstar(r)+M_{\mathrm{dm}}(r)$.  The parameter $\eta \equiv
f_{\mathrm{orb}}+f_t$ measures the orbital energy of the bulge-bulge system
when the bulges ``touch.''  $\eta$ can be decomposed into two separate
parts, with one contribution from the initial orbital energy at large radii
($f_{\mathrm{orb}}$) and the other from the subsequent energy transfer
between the stellar bulges and dark matter haloes during the merger
($f_t$); see BMQ for more details.  The values of $f$ and $\eta$ for our
equal-mass merger simulations are listed in Table~\ref{table:energy}.
These results show that, as argued in the previous paragraph, at fixed
initial orbital energy (fixed $f_{\mathrm{orb}}$), increasing the orbital
angular momentum results in a larger value of $\eta$, which is due to more
energy transfer from the bulges to the haloes (a larger positive value of
$f_t$).  This in turn leads to a more tightly bound remnant bulge with a
larger $\sigma$ and smaller $R_e$ (Fig.~\ref{fig:rperi-slopes}).
Table~\ref{table:energy} also includes the values of $\eta$ for the 0.33:1
mergers (runs P5-P8), which are quite similar to those of the equal mass
mergers.  This shows that the properties of the remnants in the projections
of the fundamental plane are not very sensitive to the mass ratio of the
merging galaxies (at least for relatively major mergers).

\begin{deluxetable}{cccccc}
\tablecaption{Energetics 
\label{table:energy}} 
\tablehead{
\colhead{\textbf{Run}\tablenotemark{a}}
& \colhead{$\mathbf{r_p}$\tablenotemark{b}} 
& \colhead{$\mathbf{f_{\mathrm{orb}}}$}
& \colhead{$\mathbf{ \eta \,\, (\equiv f_t+f_{\mathrm{orb}}})$}
& \colhead{$\mathbf{f_i}$\tablenotemark{c}}
& \colhead{$\mathbf{f_f}$}
}
\startdata
P1 & 0    & 0      & 0.225 & 0.367 & 0.521 \\         
P2 & 2.29 & 0      & 0.366 & 0.367 & 0.493 \\         
P3 & 12.5 & 0      & 0.451 & 0.367 & 0.494 \\         
P4 & 37.8 & 0      & 0.531 & 0.367 & 0.480 \\  
P5 & 0    & 0      & 0.119 & 0.295 & 0.488 \\
P6 & 2.31 & 0      & 0.317 & 0.295 & 0.457 \\
P7 & 12.5 & 0      & 0.376 & 0.295 & 0.463 \\
P8 & 20.5 & 0      & 0.445 & 0.295 & 0.460 \\
B1 & 0    & 0.145  & 0.275 & 0.367 & 0.513 \\    
B2 & 2.29 & 0.145  & 0.461 & 0.367 & 0.497 \\    
B3 & 20.8 & 0.145  & 0.539 & 0.367 & 0.488 \\    
B4 & 37.8 & 0.145  & 0.609 & 0.367 & 0.456 \\    
B5 & 57.1 & 0.145  & 0.619 & 0.367 & 0.433 \\    
B6 & 78.4 & 0.145  & 0.617 & 0.367 & 0.423 \\    
\enddata
\tablenotetext{a}{Name of run}
\tablenotetext{b}{Pericentric distance of orbit (kpc)}
\tablenotetext{c}{For 0.33:1 mergers, runs P5-P8, the value listed is
for the smaller galaxy (the larger galaxy has the same value as in the
1:1 mergers)} 
\end{deluxetable}

We can gain an analytical understanding of
Fig.~\ref{fig:rperi-slopes} and Table~\ref{table:energy} by
investigating how $\alpha$ in the $\re-\mstar$ relation varies with
the energy and structural parameters $\eta$ and $f$ in
equation~(\ref{eqn:homo1}).  Assuming that the stellar mass doubles in
equal-mass mergers (i.e. $M_{*,f}=2M_{*,i}$, a very good approximation
to the simulations) and that the stellar effective radius scales as
$\re \propto \mstar^{\alpha}$ upon merging, equation~(\ref{eqn:homo1})
relates the exponent $\alpha$ to $\eta$ and $f$:
\begin{equation}
 \alpha=1-\frac{\ln (1+\eta/4f_i)}{\ln 2}+ \frac{\ln (f_f/f_i)}{\ln 2}
  \,. \label{eqn:alpha1}
\end{equation} 
We therefore expect $\alpha=1$ if a merger is on a parabolic orbit
with negligible energy transfer ($\eta=f_{orb}+f_t=0$) and unchanged
structural parameters ($f_f=f_i$).  By contrast, if the merger orbit
initially has significant angular momentum (at distances $\gg R_i$),
the bulges will spiral inward and dynamical friction will tend to make
the orbits circular at small radii; in this case, $\eta \approx 0.5$
by the virial theorem and $\alpha \approx 0.58$ if $f_f = f_i$.  In
the actual simulations, Table~\ref{table:energy} shows there is always
nonzero energy transfer from stars to dark matter (i.e. $f_t>0$),
which decreases $\alpha$ for the head-on orbits, whereas the merged
bulge always has a larger structural parameter (i.e. $f_f/f_i > 1$),
which increases $\alpha$.  Together the variations in $\eta$ and $f$
explain the behavior and range of $\alpha$ ($\sim 0.7$ to 1.3) shown
in Fig.~\ref{fig:rperi-slopes}.

The fact that $f_f>f_i$ (by up to $\sim 40\%$) reflects the varying
central dark matter content of the remnants, though in general it is
non-trivial to quantitatively relate $f_f/f_i$ to the change in dark
matter fraction within $\re$ because equation~(\ref{eqn:f}) involves a
radial integral over the stellar density profile and dynamical mass.
However, from equation~(\ref{eqn:f}), we see that for a given
distribution of stars (fixed $\re$ and $\rho_*$), increasing the
amount of dark matter inside the bulge (i.e.  increasing $\mdyn$)
increases $f_f$, which is why $f_f > f_i$ in the simulations.

\begin{figure}
\begin{center}
\includegraphics[scale=0.45]{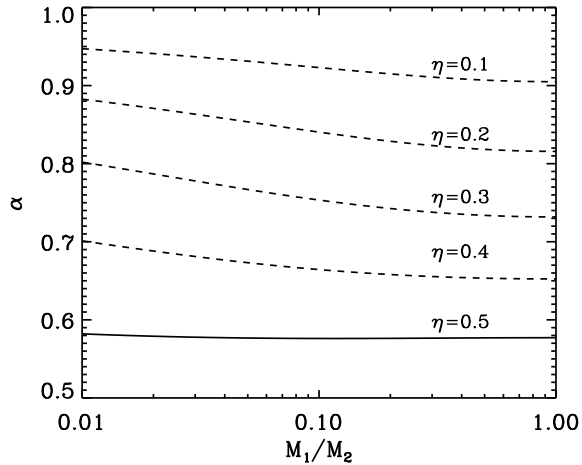}
\caption{
  Logarithmic slope $\alpha$ of the stellar mass-size relation as a
  function of the mass ratio of the merging galaxies $M_1/M_2$
  calculated from the analytical energy conservation model assuming
  $f_f=f_i$.  Head-on (parabolic) mergers correspond to $\eta\approx
  0$ while circular orbits have $\eta\approx 0.5$.  The trend of
  increasing $\alpha$ with decreasing orbital angular momentum shown
  in Fig.~3 for our simulations is reproduced in this model, although
  we do not expect the values of $\alpha$ to match exactly due to
  simplifying assumptions made in the model (e.g. $f_f=f_i$).  The
  dependence of $\alpha$ on the mass ratio is very weak, suggesting
  that the major merger results of Fig.~3 are also applicable to minor
  mergers.  This is also confirmed by the similarities between our 1:1
  and 0.33:1 simulations.}
\label{fig:alpha-m}
\end{center}
\end{figure}

The analytical energy conservation model given in
equations~(\ref{eqn:homo1}) and (\ref{eqn:alpha1}) can easily be
extended to unequal mass mergers.  Fig.~\ref{fig:alpha-m} shows the
general dependence of $\alpha$ on the mass ratio of the merging bulges
for different values of $\eta$, assuming for simplicity parabolic
orbits and $f_f=f_i$.  As argued above, $\eta$ is expected to decrease
from $\approx 0.5$ for orbits with substantial angular momentum to
$\approx 0$ for nearly parabolic head-on orbits.  Correspondingly,
$\alpha$ increases from $\approx 0.58$ to 1.  This trend of increasing
$\alpha$ with decreasing orbital circularity (or $r_{\rm peri}$) is
consistent with that shown in Fig.~3 for the simulations, although we
do not expect the values of $\alpha$ in the model and simulations to
match exactly due to simplifying assumptions made in the model (e.g.
$f_f=f_i$).  Fig.~\ref{fig:alpha-m} also shows that the dependence of
$\alpha$ on the mass ratio $M_1/M_2$ is quite weak for $0.01 \le
M_1/M_2 \le 1.0$, indicating that the results in Fig.~3 can be
extended to very minor mergers.  This analytic result is consistent
with the fact that our 0.33:1 mergers and equal mass mergers show a
similar dependence of $\alpha$ and $\beta$ on merger orbit.

\section{Discussion and Implications}
\label{sec:discuss}
\subsection{The Fundamental Plane and its Projections}

The results of Section~\ref{sec:fp} show that the \fpp\ is preserved
during dissipationless mergers for a wide range of orbits, a result of
virial equilibrium with a slight tilt due to an increasing central
dark matter fraction.  On the other hand, the results of
Section~\ref{sec:projections} show that the projections of the \fpp\ 
are quite sensitive to the amount of energy transferred from the
bulges to the dark matter haloes during the merger, which is
determined in large part by the merger orbit.  If gas-poor mergers
help build up the massive end of the red sequence, as several current
observations and models suggest, then our results indicate that the
projections of the \fpp\ -- but not necessarily the plane itself --
provide a powerful way of investigating the assembly history of
massive elliptical galaxies.

While a sample of 43 BCGs has been reported to lie on the \fpp\ 
\citep{oegerle1991}, they appear to have different \fpp\ projections
from those of lower-mass elliptical galaxies.  The correlation between
effective radius and luminosity is steeper for these BCGs ($\re
\propto L^{1.25}$) than for all early-type galaxies ($\re \propto
L^{0.6}$) and the BCG velocity dispersions lie substantially lower
than the extrapolation of the Faber-Jackson relation to higher
luminosity \citep{oegerle1991}.  Other properties of BCGs are notably
different from those of normal ellipticals as well, such as their high
incidence of secondary nuclei \citep{schneider1983}, their more
prolate shapes \citep{ryden1993}, the small dispersion in their metric
luminosities \citep{postman1995}, and the large and relatively
constant (out to $z \sim 0.4$) luminosity gap between BCGs and
second-ranked cluster galaxies \citep{loh2006}.

The results presented in this paper suggest that these differences are
all signatures of how massive elliptical galaxies were assembled.  In
particular, dissipationless merging of elliptical galaxies provides a
natural mechanism for steepening both the Faber-Jackson and $R-L$
relations \emph{and} preserving the overall \fpp, \emph{provided that
  merger orbits become preferentially more radial for the most massive
  galaxies} (Fig.~\ref{fig:rperi-slopes}).

There is, in fact, support for this possibility in the literature.
Observationally, it is well-established that BCGs are aligned with
both their host clusters and the surrounding large scale structure
(e.g. \citealt{binggeli1982,fuller1999,west2000}).  Numerical
simulations show that anisotropic merging is natural in CDM models
\citep{dubinski1998,zentner2005} and preferentially radial merging
along filaments can produce many of the observed properties of
clusters and BCGs \citep{west1994,west1995,dubinski1998}.  These lines
of reasoning favour radial mergers during BCG formation and point to a
strong connection between the characteristics of the mergers that form
BCGs and the origin of observed BCG properties (and the differences
between BCGs and normal ellipticals).  In addition, the most luminous
ellipticals in the Virgo cluster \emph{all} seem to lie on a filament
\citep{west2000}, which likely means that anisotropic merging is also
important for luminous ellipticals that are not BCGs.

According to our calculations, projections of the \fpp\ provide
independent information about the assembly history of massive
ellipticals.  Analysis of a large spectroscopic sample of massive
galaxies would provide significantly stronger constraints on changes
in the \fpp\ projections at the highest masses; we expect that the
indications of deviations seen in current samples will become more
apparent with additional data.

\subsection{Supermassive black holes}

Any departure from the canonical Faber-Jackson relation in massive
galaxies would also have significant implications for the demography
of supermassive black holes.  In particular, under the assumption that
black holes coalesce following a gas-poor merger (which is not
guaranteed; see \citealt{merritt2004c} for a review), the black hole
mass should increase roughly in proportion to the galaxy mass since
the amount of mass-energy carried away in gravitational waves is
likely to be quite small (e.g., \citealt{baker2004}).  We therefore
expect that dissipationless mergers will maintain a linear $\mbh-M_*$
relation.  Any steepening of the $L-\sigma_c$ relation for luminous
ellipticals, however, should then be reflected in a corresponding
steepening in the $\mbh-\sigma$ relation for massive black holes.
\citet{wyithe2006} has argued that just such a steepening is present
in the current sample of black hole masses, but the statistics are
rather poor and a larger sample is needed in order to make definitive
statements.  If correct, black hole masses for very luminous galaxies
based on the canonical $\mbh-\sige$ relation \citep{gebhardt2000,
  ferrarese2000, tremaine2002} would systematically under-predict the
true black hole mass (see also Lauer et al. 2006, in prep.).

\subsection{Further discussion}

Several issues can be investigated in more detail to test our suggestion
that preferentially radial merging is responsible for many of the
morphological and kinematical properties of BCGs.  One promising
possibility is to study the shapes of BCGs.  \citet{porter1991} find that
the outer regions of BCGs are generally prolate, with the inner regions
being rounder.  \citet{ryden1993} find that the distribution of BCG
ellipticities has a similar mean value to that of all ellipticals but with
a smaller dispersion and a tendency to be slightly more prolate.  For
comparison, we find that the remnants of 1:1 mergers on low angular
momentum orbits are prolate with a tendency toward triaxiality: generally
$a$:$b$:$c \sim$ 1.8:1:1 - 1.7:1.1:1.  By contrast, the runs with
significant angular momentum have $a$:$b$:$c \sim$ 1.7:1.3:1.  However, the
main difference between our equal mass and 0.33:1 mergers is that the
0.33:1 remnants are notably rounder than their 1:1 counterparts, an effect
also seen by \citet{villumsen1982}.  Unequal mass major mergers thus lead
to the same overall trends we see in the \fpp\ and its projections while
producing more spherical remnants.  Using shapes to quantitatively test the
predictions of our models will thus require detailed information about the
orbits and mass ratios of the mergers that formed BCGs.

The above considerations highlight the need to move beyond
non-cosmological major merger simulations of spherical and isotropic
galaxies that have been the focus of this paper as well as many
earlier papers.  Indeed, the relaxation and virialization of a cluster
generally involves several massive galaxies with a range of masses
\citep{dubinski1998}.  The hierarchical build-up of clusters and BCGs
also likely depends on the cluster mass (e.g.  \citealt{brough2005,
  loh2006}).  In addition, the shapes of cluster-mass dark matter
haloes in dark matter-only cosmological simulations are typically
quite prolate at formation but then become rounder (but still
triaxial) as time goes on \citep{hopkins2005, allgood2005} due to a
transition from filamentary to spherical accretion
\citep{allgood2005}.  Clearly, high resolution two-component (stars
and dark matter) simulations in a cosmological context with full
formation histories are necessary for making detailed predictions
about the properties of massive ellipticals.  We are currently working
on such simulations.  Although we expect that our basic conclusions
about the \fpp\ and its projections will remain unchanged, repeated
major and minor merging on realistic cosmological orbits will
significantly influence the shapes and kinematics of the remnant
galaxies.

As one final implication of the non-spherical nature of our merger
remnants, we note that the measured velocity dispersions can change by more
than 15\% with viewing angle for a given remnant.  For the most massive
galaxies, this could be significant: a galaxy that has $\sige \approx 350$
km/s from most viewing angles (typical for a BCG) could have $\sige > 400$
km/s in certain rare configurations (we find that the observed velocity
dispersion is $\approx \, 15 \%$ higher than the mode $\approx 3 \%$ of the
time for the remnants of relatively radial mergers).  These objects, which
are prolate merger remnants viewed down the major axis, also have smaller
effective radii than most viewing angles of remnants with the same stellar
mass.  \citet{bernardi2005} have recently reported the detection of a
population of luminous galaxies with these properties.  A simple test of
our explanation for these galaxies is to look at their projected shapes:
our calculations predict these galaxies should appear statistically rounder
than other galaxies of comparable luminosity.  This viewing angle effect
would also be a source of scatter in the $\mbh-\sige$ and Faber-Jackson
relations.

\vspace{4mm} We thank Sandy Faber, Alister Graham, Zoltan Haiman, and Tod
Lauer for useful discussions and Volker Springel for making {\tt GADGET}
publicly available.  This work used resources from NERSC, which is
supported by the US Department of Energy.  C-PM is supported in part by NSF
grant AST 0407351 and NASA grant NAG5-12173.  EQ is supported in part by
NSF grant AST 0206006, NASA grants NAG5-12043 and ATP05-54, an Alfred
P. Sloan Fellowship, and the David and Lucile Packard Foundation.

\bibliography{red_mergers}

\end{document}